\pdfoutput=1 

\documentclass[
onecolumn,
superscriptaddress,
showkeys,
preprintnumbers,
nofootinbib,
 amsmath,amssymb,
 aps,
]{revtex4-2}

\RequirePackage{fix-cm} 
\usepackage{amsmath,amssymb,amsfonts}

\usepackage{graphicx}
\usepackage{dcolumn}
\usepackage{bm}
\usepackage{bbm}
\usepackage{hyperref}
\usepackage[mathlines]{lineno}
\usepackage{booktabs}
\usepackage{multirow}
\usepackage{array}        
\usepackage{siunitx}      
\usepackage{float}
\usepackage{mathtools}
\usepackage{ulem}
\usepackage{listings}

\usepackage{fontawesome5}
\usepackage{xcolor}
\definecolor{LightGray}{gray}{0.9}

\usepackage[
]{geometry}

\usepackage{physics}
\usepackage{fancyref}
\usepackage{subcaption}
\usepackage{stfloats} 
\usepackage{placeins}
\usepackage{float}
\usepackage{caption}

\usepackage{slashed}
\usepackage{ragged2e}

\definecolor{nicered}{rgb}{0.6,0,0}
\definecolor{nicegreen}{rgb}{0.1,0.5,0.1}
\definecolor{niceblue}{rgb}{0,0.4,0.8}
\definecolor{newgreen}{rgb}{0,0.667,0}
\hypersetup{
	colorlinks=true,      
	linkcolor=black,      
	citecolor=blue,       
	filecolor=magenta,     
	urlcolor=blue
}

\usepackage{orcidlink}

\usepackage{framed}
\definecolor{shadecolor}{rgb}{0.95,0.95,0.95}
\newenvironment{claim}{\begin{shaded}\noindent\ignorespaces}{\end{shaded}}

\newcommand{\CodeURL}{https://github.com/Nape93-max/SE-BSF4DM}
\newcommand{\FeynURL}{https://cp3.irmp.ucl.ac.be/projects/feynrules/wiki/DMsimpt}

\begin{document}

\lstset{
    basicstyle=\ttfamily\footnotesize,
    keywordstyle=\color{blue},
    commentstyle=\color{gray},
    stringstyle=\color{red},
    showstringspaces=false,
    breaklines=true,
    frame=single,
    rulecolor=\color{black!30},
    backgroundcolor=\color{gray!5},
    tabsize=2,
    numbers=left,
    numberstyle=\tiny\color{gray},
    stepnumber=1,
    numbersep=5pt
}

\lstdefinelanguage{Cpp}{
    language=C++,
    morekeywords={bool, class, public, private, protected, template, namespace, using},
    sensitive=true,
    morecomment=[l]{//},
    morecomment=[s]{/*}{*/},
    morestring=[b]",
    morestring=[b]'
}

\preprint{TUM-HEP-1572/25} 
\preprint{MITP-25-077} 

\title{Manual for \texttt{SE+BSF4DM} - A micrOMEGAs package for Sommerfeld Effect and Bound State Formation in colored Dark Sectors}

\author{Mathias Becker \orcidlink{0000-0002-0203-8411}}
\email{mathias.becker@unipd.it}
\affiliation{Dipartimento di Fisica e Astronomia, Universit\`{a} degli Studi di Padova, Via Marzolo 8, 35131 Padova, Italy}
\affiliation{INFN, Sezione di Padova, Via Marzolo 8, 35131 Padova, Italy}

\author{Emanuele Copello \orcidlink{0000-0002-8404-8479}}
\email{ecopello@uni-mainz.de}
\affiliation{\text{PRISMA}$^+$ Cluster of Excellence \& Mainz Institute for Theoretical Physics, Johannes Gutenberg-Universität Mainz, 55099 Mainz, Germany}

\author{Julia Harz \orcidlink{0000-0002-8362-4083}}
\email{julia.harz@uni-mainz.de}
\affiliation{\text{PRISMA}$^+$ Cluster of Excellence \& Mainz Institute for Theoretical Physics, Johannes Gutenberg-Universität Mainz, 55099 Mainz, Germany}

\author{Martin Napetschnig \orcidlink{0000-0001-9161-9340}}
\email{martin.napetschnig@tum.de}
\affiliation{\text{PRISMA}$^+$ Cluster of Excellence \& Mainz Institute for Theoretical Physics, Johannes Gutenberg-Universität Mainz, 55099 Mainz, Germany}
\affiliation{Technical University of Munich, TUM School of Natural Sciences, Physics Department T70, 85748 Garching, Germany}

\date{\today}

\begin{abstract}
    This manual describes the usage and implementation of \href{\CodeURL}{\faGithub~\textcolor{blue}{SE+BSF4DM}}, an add-on package for \texttt{micrOMEGAs} that includes the Sommerfeld effect and bound state formation in the numerical evaluation of the dark matter relic density for QCD-colored dark sectors, applicable to any model that can be mapped onto a simplified $t$-channel framework. 
    The package seamlessly integrates these non-perturbative effects into the standard \texttt{micrOMEGAs} workflow, requiring minimal user modification. 
    This document provides a comprehensive guide to the installation, configuration, and usage of \texttt{SE+BSF4DM}, serving as a practical user guide for dark matter phenomenologists.
\end{abstract}


\maketitle

\section*{Program summary}
\textit{Program title}: \texttt{SE+BSF4DM} \\ 

\textit{Licensing provisions}: MIT \\

\textit{Programming language}: C++ \\ 

\textit{Github repository}: \href{\CodeURL}{https://github.com/Nape93-max/SE-BSF4DM} \\ 

\textit{Nature of the problem}: The accurate calculation of the dark matter relic density in models where dark sector particles interact via the strong interaction requires the inclusion of non-perturbative effects, specifically the Sommerfeld effect and bound state formation, which can alter annihilation cross sections by orders of magnitude.
While their physical significance is well-established, their implementation has remained highly model-dependent, requiring specialized expertise for each scenario. 
This has created a gap among public tools for dark matter phenomenology, as no user-friendly code seamlessly integrates these effects into the standard workflow of popular packages like \texttt{micrOMEGAs}.
Consequently, precise phenomenological studies of a broad class of well-motivated models have been hampered by the lack of an accessible and robust public code. 

\newpage

\textit{Solution method}: \textbf{SE+BSF4DM} computes the relic density by incorporating the Sommerfeld effect and bound state formation for QCD-colored dark sector particles in the (anti-)fundamental representation of $SU(3)_c$ through a modular approach into the \texttt{micrOMEGAs} framework. 
The \textbf{Sommerfeld effect} is implemented by numerically extracting the $s$-wave component of each annihilation cross section.
This $s$-wave part is then multiplied with the appropriate velocity-dependent Sommerfeld factors according to the color decomposition of the process at hand. \\
For\textbf{ bound state formation}, the code automatically identifies all possible bound states between (co-)annihilating particles and includes their weighted formation cross sections as additional annihilation channels.
This treatment makes use of implemented rates for bound state decay, ionization and transition rate between bound state levels, using analytic expressions available in the literature. 

\tableofcontents

\section{Introduction}\label{sec:intro}
This manual describes \texttt{SE+BSF4DM}, an add-on package for \texttt{micrOMEGAs} that enables the calculation of the dark matter relic density including the Sommerfeld effect and bound-state formation particles transforming in the (anti)-fundamental representation of $SU(3)_c$.
The package is designed to be user-friendly and integrates smoothly into the standard \texttt{micrOMEGAs} workflow, requiring minimal user modification while maintaining full compatibility with the code's existing functionality. \\

The physics motivation for including these non-perturbative effects is discussed in detail in the main paper Ref.~\cite{Becker2026}.
Briefly, the nature of dark matter (DM) remains one of the outstanding puzzles in particle physics and cosmology. Although weakly interacting massive particles (WIMPs) offer a simple explanation of the observed relic abundance via thermal freeze-out, direct detection and collider searches have placed increasingly strong constraints on the simplest WIMP scenarios \cite{LZ:2022lsv, PICO:2023uff, XENON:2024wpa, Arcadi:2017kky, LHCDarkMatterWorkingGroup:2018ufk, ATLAS:2023rvb}. 
These developments have motivated the exploration of more elaborate dark sectors \cite{Harris:2022vnx,cirelli2024darkmatter}, where, for example, several states participate in the thermal freeze-out and coannihilation can play a decisive role.
Such frameworks naturally accommodate heavier DM candidates \cite{Griest:1990kh, Gondolo:1990dk, Harz:2012fz, Baker:2015qna} and allow a broader range of interactions between the dark sector and the Standard Model (SM).

The increased complexity of these models has driven the development of automated tool chains that compute key DM observables, such as the relic density, direct detection rates, and collider signatures \cite{Belanger:2006is, Bringmann:2018lay, Ambrogi:2018jqj, Harz:2023llw, Palmiotto:2022rvw}. 
One of those tools is \texttt{micrOMEGAs} \cite{Belanger:2006is}, which, among several other useful functions, efficiently solves the Boltzmann equations for the time-evolution of DM in the early universe.
This solution is typically performed considering the perturbative effective annihilation cross section of DM at tree-level only.
While this approximation is sufficient for many scenarios, it breaks down in the presence of long-range interactions, which can strongly modify the annihilation rate of non-relativistic particles through non-perturbative effects \cite{Ellis:2014ipa, Ibarra:2015nca, ElHedri:2017nny, ElHedri:2018atj, Biondini:2019int}. 
This situation is particularly relevant for QCD-charged states due to the large value of $\alpha_s$. 
The \textit{Sommerfeld effect} (SE) \cite{Sommerfeld:1931qaf, Sakharov:1948plh, Beneke:2014hja, Beneke:2019qaa, Hisano:2006nn, Branahl:2019yot} can significantly enhance or suppress annihilation cross sections, while the radiative formation and decay of \textit{bound states} (BSF) \cite{vonHarling:2014kha, Ellis:2015vaa, An:2016gad, Asadi:2016ybp, Petraki:2016cnz, Liew:2016hqo} opens new, potentially dominant annihilation channels.

Despite their importance, these non-perturbative effects have remained challenging to implement in public numerical tools for the computation of the relic density. 
Until now the only code which incorporates next-to-leading order corrections including the Sommerfeld effect and can be interfaced to \texttt{micrOMEGAs} has been \texttt{DM@NLO}~\cite{Harz:2012fz,Harz:2014tma,Harz:2014gaa,Harz:2016dql,Schmiemann:2019czm,Branahl:2019yot,Harz:2022ipe,Harz:2023llw}, which is in its default implementation limited to the Minimal Supersymmetric Model (MSSM).
More recently, the tool \texttt{BSFfast}~\cite{binder2025bsffastrapidcomputationboundstate} was released, providing efficient tabulated cross sections for bound-state formation in a broad class of models.
While simplified models - particularly $t$-channel DM models (\texttt{DMSimpt} models \cite{Garny:2015wea, Giacchino:2015hvk, Mohan:2019zrk, Arina:2020udz, Arina:2020tuw, Becker:2022iso, Arina:2023msd, Jueid:2024cge, Arina:2025zpi, Biondini:2025gpg, olgoso2025darkterazfactory}) - provide benchmark scenarios with few parameters for studying cosmological and experimental constraints, the community has lacked a user-friendly, publicly available code that seamlessly integrates the SE and BSF into the standard relic density calculation specifically for this framework.

To address this gap, we developed \texttt{SE+BSF4DM}, a package for \texttt{micrOMEGAs} that enables the precise calculation of the DM relic density, including both SE and BSF for QCD colored dark sector particles.
Our tool complements existing ones by offering a fully automated, on-the-fly computation directly within the \texttt{micrOMEGAs} workflow for $t$-channel models, including the relevant color decompositions for the Sommerfeld effect.
While particularly suited for \texttt{DMSimpt} models, our package is applicable to any model where dark sector particles transform under the fundamental (or antifundamental) representation of $SU(3)_c$.
The physics results and detailed phenomenological analysis using this package are presented in the main work \cite{Becker2026}. 
This manual instead aims to give the interested user a guidance to use the code.

This manual is structured as follows: 
Section~\ref{sec:Theory_introduction} summarizes the scope and applicabilty of the package. Section~\ref{sec:install} provides instructions for installing and integrating the package into \texttt{micrOMEGAs}. Section~\ref{sec:Examples} gives detailed examples of how to configure and use the code, while Section~\ref{sec:numerics} discusses performance considerations and potential future developments.
We conclude in Section~\ref{sec:conclusions}.

\section{Scope and Applicability}\label{sec:Theory_introduction}
The package \texttt{SE+BSF4DM} \cite{Becker2026}, allows to automatically incorporate the Sommerfeld effect and bound state formation in the relic density calculation with \texttt{micrOMEGAs} for  simplified $t$-channel DM models (\texttt{DMSimpt}) as defined in Ref.~\cite{Arina:2020udz},
\begin{equation}
  \mathcal{L} = \mathcal{L}_{\rm SM} + \mathcal{L}_{\rm kin} + \mathcal{L}_F(\chi) + \mathcal{L}_F(\tilde\chi) + \mathcal{L}_S(S) + \mathcal{L}_S(\tilde S)
       + \mathcal{L}_V(V)    + \mathcal{L}_V(\tilde V) \ .\label{eq:Full_Lagrangian}
\end{equation}
The first two terms correspond to the SM Lagrangian and to the kinetic and mass terms of the dark sector, respectively. 
Fields without a tilde indicate real fields, while fields with a tilde indicate complex fields.
The remaining six terms describe the interactions that mediate $t$-channel couplings between the SM and the different DM candidates, namely Dirac and Majorana fermions ($\mathcal{L}_F$), real and complex scalars ($\mathcal{L}_S$), and real and complex vectors ($\mathcal{L}_V$), which we denote in the following by a common generic field $X$.
Explicitly one has (denoting scalar and fermionic mediators with $\varphi$ and $\psi$, respectively):\footnote{The standard \texttt{DMSimpt} implementation uses real Yukawa couplings by default, though complex values are supported.}
\begin{eqnarray}
   \mathcal{L}_F(X)& = & \Big[
           {\bf \lambda_{Q}} \bar X Q_L \varphi^\dag_{Q}
     \!+\! {\bf \lambda_{u}} \bar X u_R \varphi^\dag_{u}
     \!+\! {\bf \lambda_{d}} \bar X d_R \varphi^\dag_{d}
     \!+\! {\rm h.c.} \Big]\label{eq:S3_Lag} \ ,\\
   \mathcal{L}_S(X)& = & \Big[
          {\bf \hat\lambda_{Q}} \bar\psi_{Q} Q_L X
    \!+\! {\bf \hat\lambda_{u}} \bar\psi_{u} u_R X
    \!+\! {\bf \hat\lambda_{d}} \bar\psi_{d} d_R X
    \!+\! {\rm h.c.} \Big]\label{eq:F3_Lag} \ , \\
   \mathcal{L}_V(X)& = & \Big[
          {\bf \hat\lambda_{Q}} \bar\psi_{Q} \gamma^\mu X_\mu Q_L
    \!+\! {\bf \hat\lambda_{u}} \bar\psi_{u} \gamma^\mu X_\mu u_R
    \!+\! {\bf \hat\lambda_{d}} \bar\psi_{d} \gamma^\mu X_\mu d_R
    \!+\! {\rm h.c.} \Big]\label{eq:F3V_Lag} \ .
\end{eqnarray}
All dark sector particles are $\mathbb{Z}_2$-odd, ensuring DM stability. 
The only missing combination in the \texttt{DMSimpt} model is a (Dirac or Majorana) fermionic DM candidate $X$ with a vector mediator $Y_\mu^a$ in the fundamental representation of $SU(3)_c$.
Apart from a few exceptions \cite{Saez:2018off}, vector mediators in the fundamental representation are rarely studied, and this combination is therefore also not implemented in our package.
Moreover, renormalizable couplings in the scalar-Higgs sector are not part of the \texttt{DMSimpt} model setup.
Such couplings can be added, in which case their perturbative annihilation channels are automatically included in the relic density computation by \texttt{micrOMEGAs}.

While phenomenological studies often consider simplified cases with diagonal, flavor-universal couplings \cite{Becker:2022iso}, our implementation handles the general scenario.
To demonstrate this versatility, we provide three representative examples on \href{\CodeURL}{\faGithub~\textcolor{blue}{Github}}:

\begin{itemize}
    \item \texttt{F3SuR}: Real scalar DM with fermionic mediator coupling to right-handed up-quarks \cite{Giacchino:2015hvk, Becker2026}.
    \item \texttt{S3Muni}: Majorana DM with scalar mediator coupling to all right-handed up-type quarks \cite{Becker:2022iso}.
    \item \texttt{F3W3rd}: Complex vector DM with fermionic mediator to third-generation left-handed quarks.
\end{itemize}

By automating the inclusion of non-perturbative effects within \texttt{micrOMEGAs}, \texttt{SE+BSF4DM} significantly reduces the computational burden of precisely studying colored dark sectors, enabling a systematic exploration of the parameter space by incorporating
\begin{itemize}
  \item Sommerfeld enhancement for the velocity-independent part of the perturbative annihilation cross section,
  \item bound-state formation, including excited states, for particle--antiparticle pairs in the fundamental representation, 
\end{itemize}
within the standard thermal freeze-out framework of \texttt{micrOMEGAs}.
The Sommerfeld effect is included for QCD colored annihilations by modifying the leading term in the velocity expansion of the annihilation cross section. 
This expansion in velocity and the subsequent modification of the leading term in the velocity expansion corresponds to s-wave Sommerfeld enhancement. 
Due to this expansion in velocity, p-wave dominated processes are not captured with \texttt{SE+BSF4DM}. 
Bound state formation includes the formation, decay, ionization and transition rates between non-relativistic dark matter bound state levels (ground state as well as excited states), which enhance the total annihilation cross section.
The $t$-channel coupling $\lambda$ is assumed to be large enough to ensure chemical equilibrium in the dark sector.
If chemical equilibrium is not present in the dark sector, the used set of equations, assuming a coannihilation scenario, is no longer valid.
There is no warning given in case the chosen parameters do not allow for chemical equilibrium in the dark sector;
this must be ensured by the user.

\section{Installation and Workflow Integration}\label{sec:install}
\subsection{Prerequisites}\label{subsec:Model_definition}
Although the preparation of a CalcHEP model file is not directly related to our package, the workflow to generate the necessary model files is as follows:
\begin{enumerate}
    \item \textbf{Model customization}: The FeynRules file \texttt{dmsimpt\_v1.2.fr} provided on the \href{\FeynURL}{\texttt{DMSimpt} page} should be customized, i.e., extract only the terms relevant for the specific $t$-channel model at hand, e.g. \texttt{F3W3rd}.
    
    \item \textbf{CalcHEP output generation}: The Mathematica notebook \texttt{use-DMSimpt.nb} (also available on the \href{\FeynURL}{\texttt{DMSimpt} page} \cite{Christensen:2009jx}) is used to generate the CalcHEP output for \texttt{micrOMEGAs}.\footnote{A brief technical note: The notebook \texttt{use-DMSimpt.nb} works as expected with Mathematica\,14.0 and older versions, but throws several errors when executed with Wolfram\,14.3.}
    
    \item \textbf{Example files}: The specific FeynRules and CalcHEP files for the three benchmark models discussed in this work can be found in the \href{\CodeURL}{\faGithub~\texttt{Examples\_Tutorial/FeynRules\_CalcHEP}} directory of our GitHub repository.
\end{enumerate}

\subsection{Integrating the \texttt{SE+BSF4DM} package to \texttt{micrOMEGAs}}\label{subsec:MO_part}
\begin{figure}[h]
    \centering
    \includegraphics[width=0.8\linewidth]{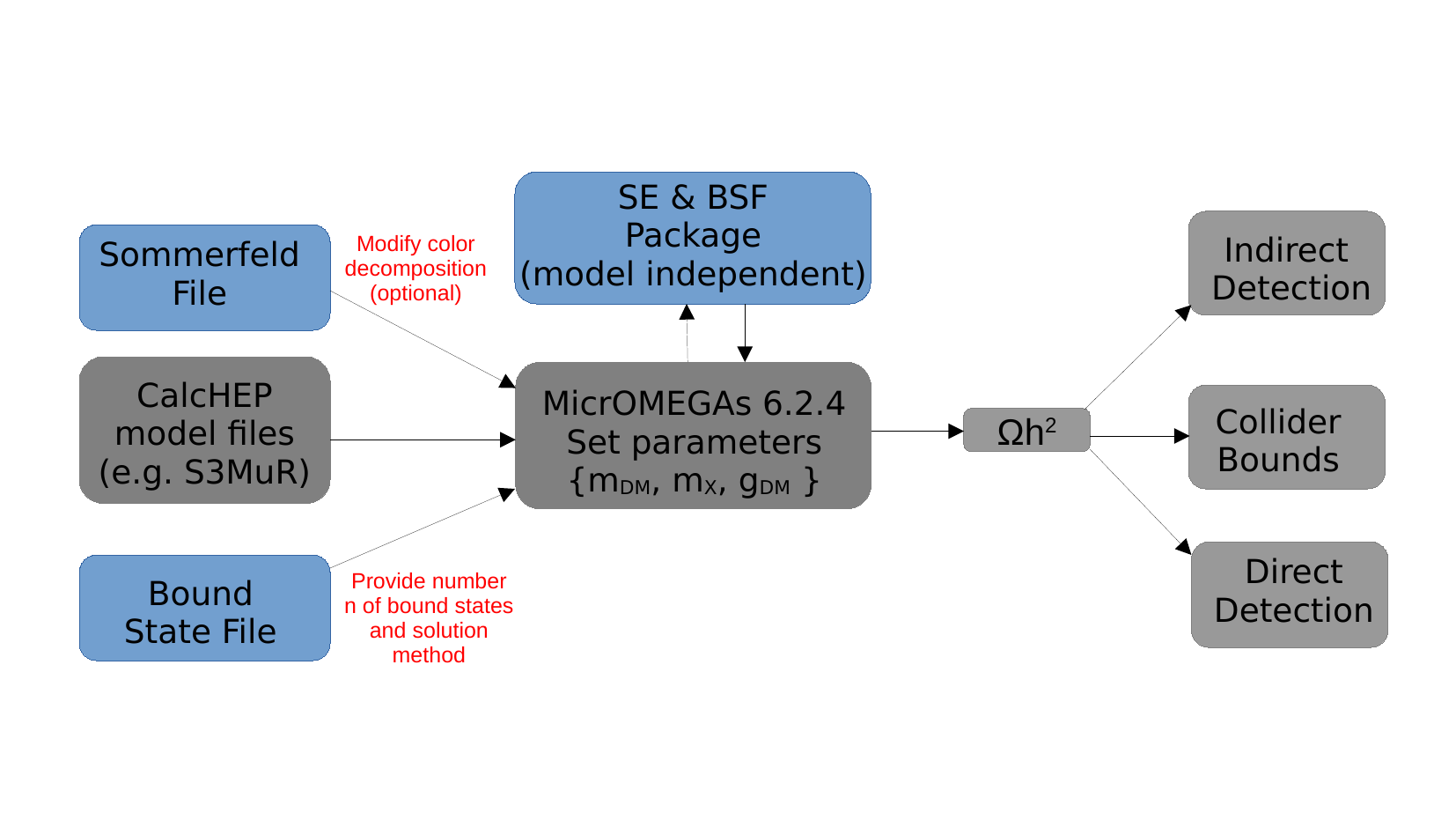}
    \caption{Schematic workflow of the \texttt{SE+BSF4DM} package.}
    \label{fig:workflow_code}
\end{figure}
To use the \texttt{SE+BSF4DM} package, \texttt{micrOMEGAs} version 6.0.~or higher has to be installed in the usual manner. 
The package is then included by copying the \texttt{copy\_into\_Packages/SE\_BSF} directory, available on our \href{\CodeURL}{\faGithub~GitHub page}, into the \texttt{micromegas\_6.2.4/Packages/} directory. 
These files contain the model-independent core functionalities of our package.
Note that this action alone does not modify the standard \texttt{micrOMEGAs} behavior; the SE and BSF effects are only activated when explicitly enabled for a specific model. To integrate our package for a certain model, the following steps have to be performed:
\begin{enumerate}
    \item Initialization of a new model by using the standard \texttt{micrOMEGAs} command \texttt{./newProject MODEL}.
    
    \item The CalcHEP model files \texttt{extlib1.mdl}, \texttt{func1.mdl}, \texttt{lgrng1.mdl}, \texttt{prtcls1.mdl} and \texttt{vars1.mdl}, which have been created as detailed in Sec.~\ref{subsec:Model_definition}, have to be transferred in to the \\ \texttt{micromegas\_6.2.4/MODEL/work/models/} directory of the newly generated model (e.g. \texttt{F3W3rd}).  
    
    \item To allow the inclusion of SE and BSF in the freshly generated model, the two model-specific files \texttt{improveCrossSection\_Sommerfeld.cpp} and \texttt{BoundStateFormation.cpp}, contained in the \texttt{copy\_into\_MODEL\_lib/} directory in our \href{\CodeURL}{\faGithub~ GitHub repository},  have to be copied into the \texttt{micromegas\_6.2.4/MODEL/lib/} folder.
\end{enumerate}
Finally, these model-specific files have to be configured as detailed in the following section.

\section{Usage and Examples}\label{sec:Examples}
\subsection{General Set-Up and Perturbative Annihilations}

Before activating any non-perturbative effects, the first step is to integrate the model-specific files \texttt{lib/improveCrossSection\_Sommerfeld.cpp} and \texttt{lib/BoundStateFormation.cpp} into the \texttt{micrOMEGAs} build. This is done by inserting them in the \texttt{main.cpp} file of the model directory via a preprocessor directive, as shown in code example~\ref{listing:include_package}. \\

\begin{minipage}{\textwidth}
\begin{lstlisting}[
  frame=trbl,
  framerule=0.4pt,
  framesep=2mm,
  xleftmargin=15pt,  % Critical: makes space for line numbers
  xrightmargin=2mm,
  backgroundcolor=\color{gray!5},
  basicstyle=\small\ttfamily,
  language=Cpp,
  numbers=left,
  numbersep=1pt,
  numberstyle=\tiny\color{gray}\hfill,
  breaklines=true
]
/* #include"../include/micromegas.h"
#include"../include/micromegas_aux.h"
If our SE+BSF4DM is used, these two files are included directly in 
"lib/improveCrossSection_Sommerfeld.cpp" and "lib/BoundStateFormation.cpp" */
#include"lib/pmodel.h"
#include"lib/improveCrossSection_Sommerfeld.cpp"
#include"lib/BoundStateFormation.cpp"
\end{lstlisting}
\captionof{lstlisting}{Files to be included in the \texttt{main.cpp} file in \texttt{micrOMEGAs}.}\label{listing:include_package}
\end{minipage}\\

As long as the corresponding flags, \texttt{somm\_flag} in \texttt{lib/improveCrossSection\_Sommerfeld.cpp} and \texttt{bsf\_scenario} in \texttt{lib/BoundStateFormation.cpp}, are set to zero, the relic density is computed using the standard perturbative cross sections. \\

The implementation of non-perturbative effects leverages and extends \texttt{micrOMEGAs}' existing infrastructure while maintaining computational efficiency. 
The calculation of the dark matter relic density in \texttt{micrOMEGAs} \cite{Belanger:2010pz} automatically accounts for coannihilation processes through the effective annihilation cross section \cite{Griest:1990kh}
\begin{align}
    \langle\sigma_{\text{eff}} v_{\text{rel}}\rangle&=\sum_{\alpha \beta}\langle\sigma_{\alpha \beta }v_{\alpha \beta}\rangle \dfrac{Y_\alpha^{\text{eq}}}{\tilde{Y}^{\text{eq}}}\dfrac{Y_\beta^{\text{eq}}}{\tilde{Y}^{\text{eq}}}
    \label{eq:effective_sigmav_coannih}, \\
    \tilde{Y}^{\text{eq}}&\equiv \sum_\alpha Y_\alpha^{\text{eq}},
\end{align} 
where $Y_\alpha^{\text{eq}}$ denotes the equilibrium yield of dark sector species $\alpha$ (distinct from the $t$-channel mediators introduced previously), and $\langle\sigma_{\alpha\beta} v_{\alpha \beta}\rangle$ represents the thermally averaged annihilation cross section for $\alpha + \beta \rightarrow \text{SM}$ processes.
In \texttt{micrOMEGAs}, the routine used to calculate the dark matter relic abundance, assuming a coannihilation scenario with an effective cross section given by Eq.~\eqref{eq:effective_sigmav_coannih}, is \texttt{darkOmega}.
The averaged annihilation cross section in \texttt{micrOMEGAs} is calculated with the function \texttt{vSigmaA}.

In order to include the Sommerfeld effect and bound-state formation the effective cross section $\expval{\sigma_{\alpha \beta} v_{\alpha \beta}}$ is replaced by
\begin{claim}
\begin{equation}
\expval{\sigma_{\alpha \beta} v_{\alpha \beta}} = \expval{\mathcal{S} \left( \sigma_{\alpha \beta} v_{\alpha \beta} \right) }_\text{ann} + \sum_{\substack{
    \alpha + \beta \rightarrow \mathcal{B}_i \\
    \rightarrow \text{SM SM}
}} \expval{\sigma_{\text{BSF}, i} v_\text{rel}} \underbrace{\left(1 - \sum_j \left( M^{-1}\right)_{ij} \frac{\expval{\Gamma^j_\text{ion}}}{\expval{\Gamma^j}} \right)}_{=:\mathcal{W}_i}. \label{eq:BSF_master_equation}
\end{equation}
\end{claim}
The individual terms appearing in Eq.~\eqref{eq:BSF_master_equation} are detailed and explained in the following two subsections. 

\subsection{Sommerfeld Effect}\label{subsec:code_description_SE}
\subsubsection{Conceptional integration into \texttt{micrOMEGAs}}\label{subsubsec:SE_conceptual_integration}
The first term in Eq.~\eqref{eq:BSF_master_equation} represents the Sommerfeld-corrected annihilation cross section.
In the \texttt{micrOMEGAs}, this term can be consistently included in the computation of the effective annihilation cross section, accessible via \texttt{SigmaA}, and of the relic density computation when using the \texttt{darkOmega} or \texttt{darkOmegaExt} routines.
The symbol $\mathcal{S}$ represents the Taylor expansion in velocity, the color decomposition and the subsequent Sommerfeld enhancement/suppression of the unaveraged cross section. 
These steps are performed using the \texttt{micrOMEGAs} built-in function \texttt{improveCrossSection}.
The function \texttt{improveCrossSection} allows to modify the unaveraged cross section, meaning that, for every colored annihilation process, the conventional perturbative cross section is substituted with the Sommerfeld enhanced unaveraged cross section. 
As it is a substitution and not an addition, double counting of processes is avoided. 

Focusing on s-wave contribution, we truncate the series expansion of the cross section in partial waves at zeroth order \cite{Cassel:2009wt, ElHedri:2016onc, Becker2026} 
\begin{align}
    \mathcal{S}\left( \sigma_{\alpha \beta} v_\text{rel} \right) &= \sum_{s} \sum_{[\mathbf{R}]} c^{[\textbf{R}]}_{s} S_0^{[\textbf{R}]}(\zeta)\,\sigma ^{(0, s)}_{\alpha \beta} v_\text{rel} + \mathcal{O}(v_\text{rel}^2),\label{eq:Sommerfeld_corrected_XS}
\end{align}
where
\begin{align}
    S_0^{[\textbf{R}]}(\zeta) &\equiv S_0 \left(k_{[\textbf{R}]}\frac{\alpha_s}{v_{\text{rel}}}\right)\, ,\label{eq:somfact_color}
\end{align}
is the representation-dependent Sommerfeld factor for the s-wave contribution, which is given by
\begin{align}
    S_0(\zeta) &= \frac{2\pi\zeta}{1-e^{-2\pi\zeta}} \, .\label{eq:S0_coulomb}
\end{align}
In Eq.~\eqref{eq:Sommerfeld_corrected_XS}, the velocity expansion extracts the s-wave component in the $v_\text{rel} \rightarrow 0$ limit ($\sigma^\text{s-wave}\propto \frac{1}{v_\text{rel}} + \mathcal{O} (v_\text{rel})$), as detailed in Appendix A in our main paper \cite{Becker2026}. 
While \texttt{micrOMEGAs} contains built-in Sommerfeld functions optimized for massive force mediators \cite{Alguero:2023zol}, our package employs the more efficient analytic expressions \eqref{eq:S0_coulomb} for the QCD Coulomb potential featuring massless force mediators (gluons). 
The coefficients $c^{[\textbf{R}]}_{s}$ (with $\sum_s c^{[\textbf{R}]}_{s} = 1$) decompose the cross section by final-state spins, as the color decomposition depends on the spin configuration \cite{ElHedri:2016onc}, as detailed in Ref.~\cite{Becker2026} and exemplarily shown in Eqs.~\eqref{eq:color_decomp_pure8}-\eqref{eq:color_decomp_qq}. 
The running of the QCD coupling is evaluated at scale $\alpha_s(Q = \mu v_\text{rel})$ employing \texttt{micrOMEGAs}' built-in function \texttt{alphaQCD(Q)} for the strong running coupling, where $\mu$ is the reduced mass of the annihilating pair. 
The color coefficients $k_{[\textbf{R}]}$ in Eq.\eqref{eq:somfact_color} are real numbers and depend on the initial state's color representation. 
If $k_{[\textbf{R}]}>0$, the color potential is attractive and the Sommerfeld effect results in an enhancement of the cross section, while if $k_{[\textbf{R}]}<0$, the color potential is repulsive and the Sommerfeld effect results in a suppression of the cross section.

\subsubsection{Default Sommerfeld implementation}\label{subsubsec:default_color_decomp}
The default implementation, employs the color decomposition valid for \texttt{DMSimpt} models under the assumption $\lambda \ll g_s$,\footnote{In the opposite regime $\lambda \gg g_s$, coannihilations become less important, as the processes $X X \rightarrow q q$ dominate.} resulting in the following color decompositions \cite{Giacchino:2015hvk, ElHedri:2016onc, Becker2026}

\begin{align}
    \mathcal{S}\sigma_{\overline{Y} Y \rightarrow g \mathcal{A}} &= S^{[\textbf{8}]}_0 \sigma_{\overline{Y} Y  \rightarrow g \mathcal{A}},\label{eq:color_decomp_pure8} \\
    \mathcal{S}\sigma_{\overline{Y} Y  \rightarrow \mathcal{B} \mathcal{C}} &= S^{[\textbf{1}]}_0 \sigma_{\overline{Y} Y \rightarrow \mathcal{B} \mathcal{C}},\label{eq:color_decomp_pure1} \\
    \mathcal{S}\sigma_{\overline{Y} Y  \rightarrow g g} &= \left(\frac{2}{7} S^{[\textbf{1}]}_0 + \frac{5}{7} S^{[\textbf{8}]}_0 \right) \sigma_{X X^\dagger \rightarrow g g},\label{eq:color_decomp_gg} \\
    \mathcal{S}\sigma_{\overline{Y}_i Y_j \rightarrow \overline{q}_i q_j} &= \sigma_{\overline{Y}_i Y_j \rightarrow \overline{q}_i q_j} \times  
    \begin{cases}
        S^{[\textbf{8}]}_0, \text{$Y$ fermionic \& } i=j  \\ 
        \frac{1}{9} S^{[\textbf{1}]}_0 + \frac{8}{9} S^{[\textbf{8}]}_0, \text{ else}
    \end{cases} ,\label{eq:color_decomp_qqbar} \\
    \mathcal{S}\sigma_{Y_i Y_j \rightarrow q_i q_j} &= \sigma_{Y_i Y_j \rightarrow q_i q_j} \times
    \begin{cases}
        S^{[\textbf{6}]}_0, \text{$Y$ scalar \& } i=j  \\ 
        \frac{1}{3} S^{[\overline{\textbf{3}}]}_0 + \frac{2}{3} S^{[\textbf{6}]}_0, \text{ else}
    \end{cases}.\label{eq:color_decomp_qq}
\end{align}
Here, $\mathcal{A}$ in Eq.~\eqref{eq:color_decomp_pure8} denotes color-neutral SM vector bosons $\{A, W^\pm, Z\}$, while $\mathcal{B}$ and $\mathcal{C}$ in Eq.~\eqref{eq:color_decomp_pure1} represent any color-neutral SM particles $\{A, W^\pm, Z, h, \ell\}$. 
The indices $i, j$ in Eqns.~\eqref{eq:color_decomp_qqbar} and \eqref{eq:color_decomp_qq} label quark and mediator flavors. 
A more detailed discussion on the physics of these decompositions (related to the symmetry of the overall wavefunction) is given in Ref.~\cite{Becker2026}. 
In all three available model files on \href{\CodeURL}{\faGithub~\textcolor{blue}{Github}}, the color decompositions \eqref{eq:color_decomp_pure8}-\eqref{eq:color_decomp_qq} are employed in the \texttt{improveCrossSectionSommerfeld.cpp} files.
The specific coefficients \texttt{kQfac} correspond to the coefficients $c^{[\textbf{R}]}_{s}$, introduced around Eq.~\eqref{eq:Sommerfeld_corrected_XS}. 
Code example~\ref{listing:color_decomp} shows the key implementation, with the coefficients \texttt{kQfac1}, \texttt{kQfac3}, \texttt{kQfac6}, and \texttt{kQfac8} corresponding to the representations $\mathbf{1}$, $\overline{\mathbf{3}}$, $\mathbf{6}$ and $\mathbf{8}$, respectively (if unspecified, \texttt{kQfac = 0}). For example, line 11 in code example~\ref{listing:color_decomp} describes the process $\overline{Y} Y \rightarrow gg$ (see Eq.~\eqref{eq:color_decomp_gg}), while line 32 describes the process $Y_i Y_j \rightarrow q_i q_j, \, i \neq j$  (see Eq.~\eqref{eq:color_decomp_qq}) for different flavors of mediators.  

These steps are implemented using \texttt{micrOMEGAs}' built-in \texttt{improveCrossSection} method, with the resulting improved cross section undergoing standard thermal averaging.
The implementation of the color decomposition in the \texttt{improveCrossSection} function is shown below in code example \ref{listing:color_decomp}.
The core computation takes place in the function \texttt{SommerfeldFactor\_BSMmodel}, where the color decomposition is handled through conditional \texttt{if...else} blocks that identify specific initial and final states.

\begin{minipage}{\textwidth}
\begin{lstlisting}[
  frame=trbl,
  framerule=0.4pt,
  framesep=2.5mm,
  xleftmargin=2pt,  % Critical: makes space for line numbers
  xrightmargin=2mm,
  backgroundcolor=\color{gray!5},
  basicstyle=\small\ttfamily,
  language=Cpp,
  numbers=left,
  numbersep=0.1pt,
  numberstyle=\tiny\color{gray}\hfill,
  breaklines=true
]
    // *** BEGIN IF BLOCK OF COLOR DECOMPOSITION (CAN BE MODIFIED OPTIONALLY) ***
  if((c1==3&&c2==-3)||(c1==-3&&c2==3)){ // Y Ybar process
        if((c3!=8&&c4==8)||(c3==8&&c4!=8)){ //g + Z/\gamma. 
    // This is purely adjoint for all partial waves. 
      kQfac8=1.;
      } 
    if(c3==1&&c4==1){ // most frequent case: Both final states are colour singlets
      kQfac1=1.; //for all partial waves 
      }
    if(c3==8&&c4==8){ //gg final state
      kQfac1=2./7.; kQfac8=5./7.; //gg channel
      } 
    if((c3==3&&c4==-3)||(c3==-3&&c4==3)){ //q qbar final state
      /* This is the tricky channel, as elaborated in our publication.
      Interference terms in the color decomposition are neglected   */
      if((s1==1 && s2==1)&&(n3 == -n4)){ 
        // fermionic mediators and identical quark flavors in the final state
        kQfac8=1.; // for all partial waves in the case \lambda << g_s
      }
      else{ // different quark flavors in the final state or scalar mediators
        kQfac1=1./9.; kQfac8=8./9.; 
        //for all partial waves in the case \lambda << g_s
        }
      }
    }
  if((c1==3&&c2==3)||(c1==-3&&c2==-3)){ // q_i q_j or q_i_bar, q_j_bar final state
    if((s1==0 && s2==0)&&(n3==n4)) { 
      // Scalar mediators and equal quark flavors in the final state
      kQfac6=1.; 
    } 
    else {
      kQfac3=1./3.; kQfac6=2./3.; 
      }  
    }
    // *** END IF BLOCK OF COLOR DECOMPOSITION ***
\end{lstlisting}
\captionof{lstlisting}{Color decomposition implementation for Sommerfeld enhancement in \texttt{lib/improveCrossSection\-Sommerfeld.cpp}.
The complete block handles all relevant initial/final state combinations for \texttt{DMSimpt} models.}\label{listing:color_decomp}
\end{minipage} \\ 

\subsubsection{Configuration}

To activate the Sommerfeld enhancement only, users must set \texttt{somm\_flag = 1} in \\ \texttt{lib/improveCrossSection\_Sommerfeld.cpp}. The implementation then calculates velocity-dependent enhancement factors for annihilation channels by decomposing the initial state into irreducible color representations. 


\subsubsection{Possible extensions by the user}\label{subsubsec:SE_potential_extensions}
Users may modify the coefficients $c^{[\textbf{R}]}_{s}$ for scenarios such as $\lambda \gg g_s$, with implementation details provided in Sec.~\ref{subsubsec:SE_conceptual_integration} and code example~\ref{listing:color_decomp}. 
Users are also able to implement Sommerfeld corrections to higher partial waves or loop-improved cross sections by setting \texttt{somm\_flag = 2} and modifying the user-defined block at the end of the \texttt{improveCrossSection} function, maintaining compatibility with \texttt{micrOMEGAs}' original purpose for this function, namely allowing the user to modify the unaveraged cross section at will. 

The Sommerfeld effect is computed by numerically extracting the leading order term in the Taylor expansion of the \texttt{micrOMEGAs}-internally computed $\sigma_\text{ann}v_\text{rel}(v_\text{rel})$, which coincides with the s-wave contribution in the limit $v_\text{rel} \rightarrow 0$.\footnote{This approach can not be extended to the p-wave, due to inseparable mixing between s- and p-wave contributions in the $v^2_\text{rel}$ term in the expansion.} 
Depending on the color charge of the particles in the initial and final states, the appropriate color decomposition (which can be straightforwardly modified by the user) is employed, as described in Sec.~\ref{subsubsec:default_color_decomp} and code example~\ref{listing:color_decomp}. 

The Sommerfeld effect for colored annihilating particles is implemented in three steps and only optionally requires modifications by the user:
\begin{enumerate}
    \item Taylor expansion in velocity to extract the $v_\text{rel}^0$ contribution.
    This step is fully automated for the leading $v_\text{rel}^0$ and does not require user's modification.
    If one whishes to go beyond this order, the user has to implement their own expansion using the function \texttt{improveCrossSection}.
    \item Color decomposition for initial states, i.e. setting the coefficients $c^{[\textbf{R}]}_{s}$, introduced around Eq.~\eqref{eq:Sommerfeld_corrected_XS}. 
    If the user adopts the default color decompositions~\eqref{eq:color_decomp_pure8}-\eqref{eq:color_decomp_qq}, nothing more has to be done by the user. 
    If the user whishes to use a different decomposition, they have to give different values to the variables \texttt{kQfac}, corresponding to the coefficients $c^{[\textbf{R}]}_{s}$, in the \texttt{improveCrossSection} function.
    \item Define enhancement/suppression factors according to Eqns.~\eqref{eq:somfact_color}-\eqref{eq:S0_coulomb}. 
    This is fully automated and does not require any modification by the user.
\end{enumerate}
\subsection{Bound State Formation}\label{subsec:code_description_BSF}

\subsubsection{Conceptional integration into \texttt{micrOMEGAs}}\label{subsubsec:BSF_conceptual}
\begin{figure}[h]
    \centering
    \includegraphics[width=.85\linewidth]{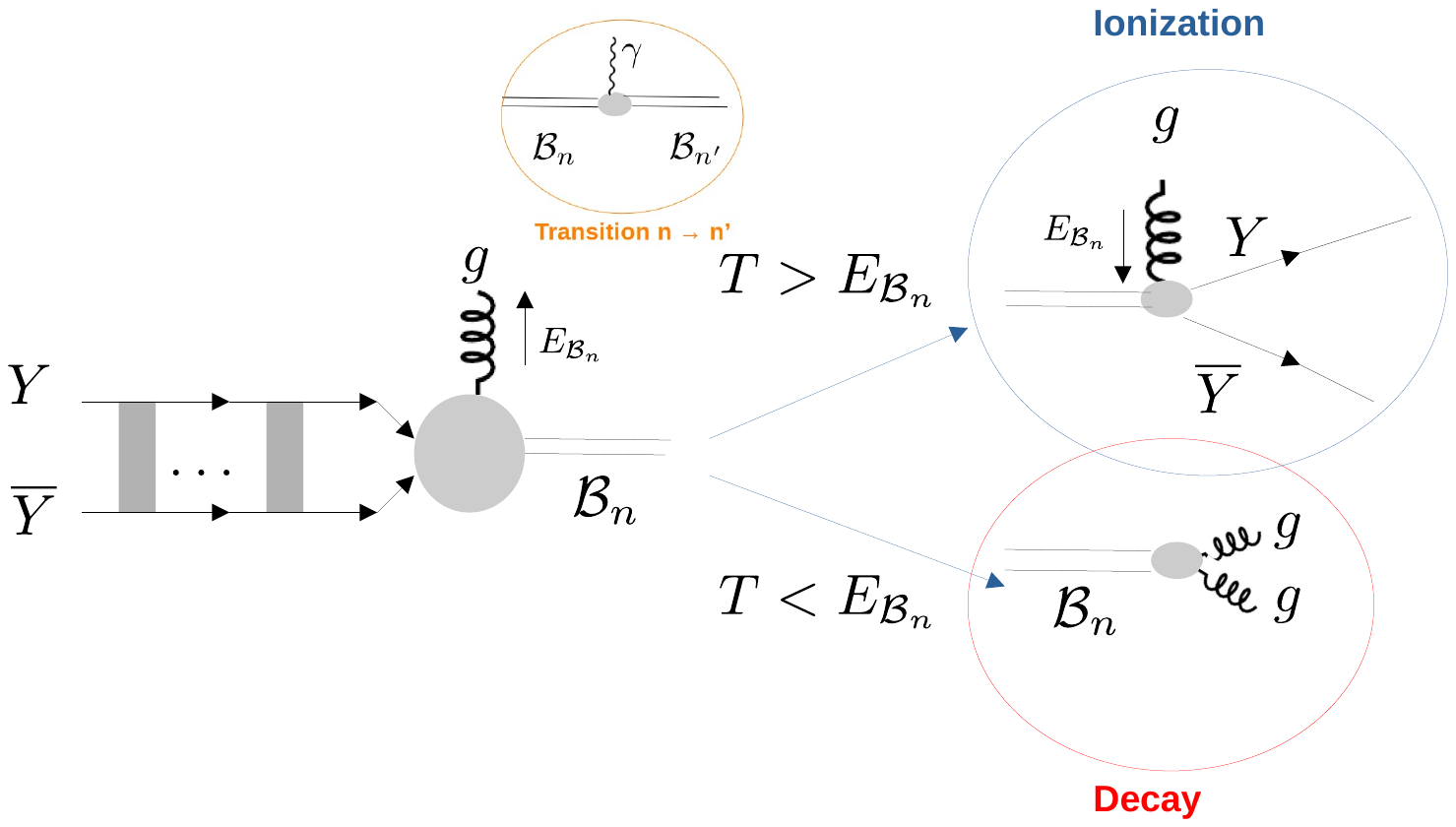}
    \caption{Schematic BSF process with ionization, decay, and transitions.}
    \label{fig:sketch_BSF}
\end{figure}

The second term in Eq.~\eqref{eq:BSF_master_equation} accounts for bound-state formation (BSF), where annihilating particles in a $\mathbf{3} \otimes \mathbf{\overline{3}}$ color configuration form hydrogen-like metastable bound states $\mathcal{B}_i$ with quantum numbers $i = \{n, \ell, m, s\}$ that subsequently decay into Standard Model particles. 
This represents an additional annihilation channel beyond the Sommerfeld-modified perturbative processes.
In \texttt{micrOMEGAs}, the second term in Eq.~\eqref{eq:BSF_master_equation} can be used utilizing the \texttt{darkOmegaExt} routine.
The derivation of the effective cross section in Eq.~\eqref{eq:BSF_master_equation} follows Refs.~\cite{Binder:2021vfo, Garny:2021qsr}. 

The weight factor $\mathcal{W}_i$ in Eq.~\eqref{eq:BSF_master_equation},
\begin{align}
        \mathcal{W}_i &= 1 - \sum_j \left( M^{-1}\right)_{ij} \frac{\expval{\Gamma^j_\text{ion}}}{\expval{\Gamma^j}},
\end{align}
accounts for ionization by thermal particles ($\langle \Gamma^i_\text{ion} \rangle$), decay into gluons ($\langle \Gamma^i_\text{dec} \rangle$), and transitions between bound-state levels ($\langle \Gamma^{i\rightarrow j}_\text{trans} \rangle$) mediated by SM photons.\footnote{In \texttt{DMSimpt} models, mediators carry electric charge by construction.
The strong interaction cannot mediate transitions.
If mediators $Y$ are charged under $SU(2)_L$, there are additional $\mathbf{3}_{SU(2)_L} \rightarrow \mathbf{1}_{SU(2)_L}$ transitions present which are not taken into account at the moment.
}
The various processes are illustrated in Figure~\ref{fig:sketch_BSF}. 
A more detailed discussion about the matrix $M$ and the appearing rates $\Gamma_i$ is given in Ref.~\cite{Becker2026}. The matrix $M$ has dimension $\left((2s + 1) n^2\right)^2$, when including states up to principal quantum number $n$. 
This leads to a $\mathcal{O}(n^4)$ scaling that significantly impacts computational performance (see Section~\ref{subsec:Performance}).
To ensure the validity of the Coulomb potential, we require binding energies $E_{\mathcal{B}_i} = \frac{\mu\left(\alpha_b^{(i)}\right)^2}{2n^2} \overset{!}{>} \Lambda_\text{QCD}$, yielding $n \lesssim 20$ for $\mathcal{O}(1\ \text{TeV})$ masses, where $\Lambda_\text{QCD}$ is the QCD confinement scale and $\alpha_b^{(i)}$ is the strong coupling constant evaluated at the Bohr momentum of bound state $i$. 
Our code does not automatically check for the validity of $E_{\mathcal{B}_i} = \frac{\mu\left(\alpha_b^{(i)}\right)^2}{2n^2} \overset{!}{>} \Lambda_\text{QCD}$.
This does not constitute a problem in practice, as in coannihilations excited states with $n>20$, for which the hierarchy $E_{\mathcal{B}_i} = \frac{\mu\left(\alpha_b^{(i)}\right)^2}{2n^2} \overset{!}{>} \Lambda_\text{QCD}$ could be violated, are shown to be irrelevant \cite{Garny:2021qsr}.

We implement the rates from Ref.~\cite{Binder:2023ckj} for $\sigma_{\text{BSF}, i} v_\text{rel}$, computing thermal averages using \texttt{micrOMEGAs}' Simpson integration.\footnote{Ref.~\cite{Beneke:2024nxh} provides equivalent updated expressions. 
Unitarity issues raised in Refs.~\cite{Binder:2023ckj, Beneke:2024nxh} are negligible for $n \lesssim 20$.} 
The full BSF contribution is added to the effective cross section via \texttt{micrOMEGAs}' \texttt{darkOmegaExt} routine. 
Users may select from the four computational schemes \textbf{No Transition Limit (1)}, \textbf{Efficient Transition Limit (2)}, \textbf{Ionization Equilibrium (3)} and \textbf{Full Solution (4)}, which are further explained in Refs.~\cite{Garny:2021qsr, Becker2026}, by setting the flag \texttt{bsf\_scenario} to \texttt{1}, \texttt{2}, \texttt{3} or \texttt{4}, respectively.

The package computes then the total effective cross section by summing \texttt{micrOMEGAs}' standard annihilation channels (potentially including Sommerfeld enhancement) with all relevant bound-state formation processes.
Ionization and decay rates are taken from Ref.~\cite{Garny:2021qsr}, while BSF and transition rates are taken from Ref.~\cite{Binder:2023ckj}.  
As the running coupling $\alpha_s$ that enters in the BSF and decay rates has to be evaluated at different scales~\cite{Petraki:2016cnz, Harz:2018csl}, these values need to be self-consistently solved numerically, e.g. the coupling at the scale of the Bohr momentum $\alpha^{(i)}_b = \alpha_s\left(Q = \mu \alpha^{(i)}_b\right)$. 
The calculation employs \texttt{micrOMEGAs}' built-in function \texttt{alphaQCD(Q)} for the strong running coupling and obtains $\alpha^{(i)}_b$ numerically via the Newton-Raphson method \cite{Scherer_comp_phys}.
The \texttt{simpsonArg} routine is used for thermal averaging the BSF rates, which is also used for other thermal averages in \texttt{micrOMEGAs}. 
For the \textbf{Full matrix} solution, the matrix $M$ containing all the transition rates is diagonalized using \texttt{micrOMEGAs}' internal \texttt{rDiagonalA} diagonalization routine.

\subsubsection{Configuration}
In \texttt{lib/BoundStateFormation.cpp}, users have to specify the treatment of bound states via the \texttt{bsf\_scenario} flag and the number of excited states via \texttt{num\_excited\_states}, as shown in code example~\ref{listing:bsf_settings}. 
The available scenarios, detailed in Section~\ref{subsubsec:BSF_conceptual} and in Ref.~\cite{Becker2026}, are the computationally efficient \textbf{No Transition} limit (1), the \textbf{Efficient Transition} limit (2), the \textbf{Ionization Equilibrium} (3), and the \textbf{Full Matrix} solution (4). 
In simplified $t$-channel models, scenario 1 provides an excellent approximation to the full solution. 
Choosing one of these options leads to the standard relic density calculation in \texttt{main.cpp} being replaced by the extended version that includes bound-state effects, as shown in code example~\ref{listing:bsf_main_darkOmega}.
The \texttt{darkOmegaExt} interface also allows implementation of custom cross section enhancements by replacing the final two arguments of \texttt{darkOmegaExt} with user-defined functions, maintaining full compatibility with the original functionality.
In our implementation, the method \texttt{darkOmegaExt} includes bound state formation by adding to the perturbative (co-)annihilations (optionally including the Sommerfeld effect, depending on the flag that the user sets) the effective bound state formation cross section, such that \texttt{darkOmegaExt} implements the master equation~\eqref{eq:BSF_master_equation}.
For comparison, the \texttt{darkOmega} routine calculates the dark matter relic density using only the perturbative (co-)annihilations, again with or without the Sommerfeld effect.
Both methods solve the Boltzmann equation for the dark matter yield numerically to a temperature of \texttt{Tend = 0.001}\,GeV, which is the default value implemented in \texttt{micrOMEGAs} that we leave unchanged.
This cutoff captures all bound state dynamics safely for all mediator masses down to around $\mathcal{O}(10)$\,GeV. 

\begin{minipage}{\textwidth}
\begin{lstlisting}[
  frame=trbl,
  framerule=0.4pt,
  framesep=2.5mm,
  xleftmargin=2pt,  % Critical: makes space for line numbers
  xrightmargin=2mm,
  backgroundcolor=\color{gray!5},
  basicstyle=\small\ttfamily,
  language=Cpp,
  numbers=left,
  numbersep=0.1pt,
  numberstyle=\tiny\color{gray}\hfill,
  breaklines=true
]
#include"../../Packages/SE_BSF/SE_BSF_header.h"
#include"../../Packages/SE_BSF/SE_BSF_functions.cpp"
/* In this file, the user has to supply some details of the model for BSF. The following information needs to be provided:
 1) The BSF scenario/limit
 2) The number of excited states to be considered */
int bsf_scenario = 1; //Flag for BSF (0 = no BSF, 1 = no transition limit, 2 = efficient transition limit, 3 = ionization equilibrium, 4 = Full matrix solution)
int num_excited_states = 1; //Number of n states included in the calculation (n = 0 -> no bound states, n = 1 -> ground state etc.)
// *** END USER DEFINITION ***
\end{lstlisting}
\captionof{lstlisting}{User defined settings in the \texttt{lib/BoundStateFormation.cpp} file for activating and controlling bound-state effects.}\label{listing:bsf_settings}
\end{minipage}\\


\begin{minipage}{\textwidth}
\begin{lstlisting}[
  frame=trbl,
  framerule=0.4pt,
  framesep=2.5mm,
  xleftmargin=2pt,  % Critical: makes space for line numbers
  xrightmargin=2mm,
  backgroundcolor=\color{gray!5},
  basicstyle=\small\ttfamily,
  language=Cpp,
  numbers=left,
  numbersep=0.1pt,
  numberstyle=\tiny\color{gray}\hfill,
  breaklines=true
]
if(Ncdm==1) 
  {  double Xf;
     //Omega=darkOmega(&Xf,fast,Beps,&err); //Conventinal MO calculation without BSF
     Omega = darkOmegaExt(&Xf, BSF_XS_A, BSF_XS_S); // MO calculation with BSF
     printf("Xf=%.2e Omega=%.2e\n",Xf,Omega);
     if(Omega>0)printChannels(Xf,cut,Beps,1,stdout);}
\end{lstlisting}
\captionof{lstlisting}{Usage of \texttt{darkOmegaExt} in the \texttt{main.cpp} file to include (excited) bound state effects.}\label{listing:bsf_main_darkOmega}
\end{minipage}\\

\subsubsection{Comparison of bound-state solution methods}
\begin{figure}[!h]
    \centering
    \includegraphics[scale=0.5]{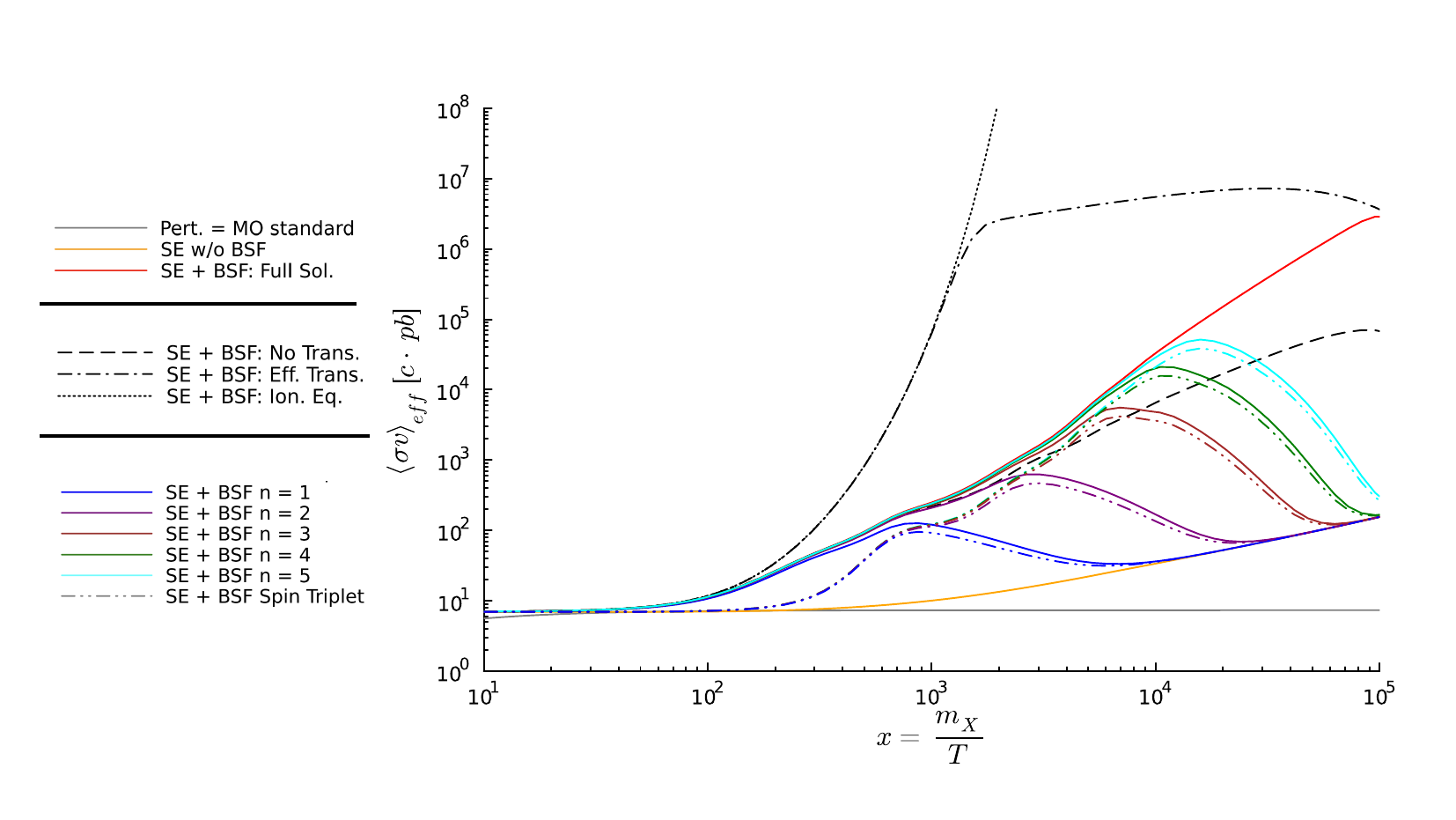}
    \caption{Comparison of effective cross sections for different bound-state treatment methods in the \texttt{F3SuR} model with mass degenerate dark sector particles $m_X = m_Y = 1$\,TeV and a $t-$channel coupling $\lambda = 0.01$. 
    All black lines are computed including $n = 15$ bound states according to the four methods discussed in the main text.
    Colors indicate contributions from different principal quantum numbers $n$ using the full matrix solution.}
    \label{fig:Cross_Sections}
\end{figure}
\begin{figure}[h]
    \centering
    \includegraphics[scale=0.5]{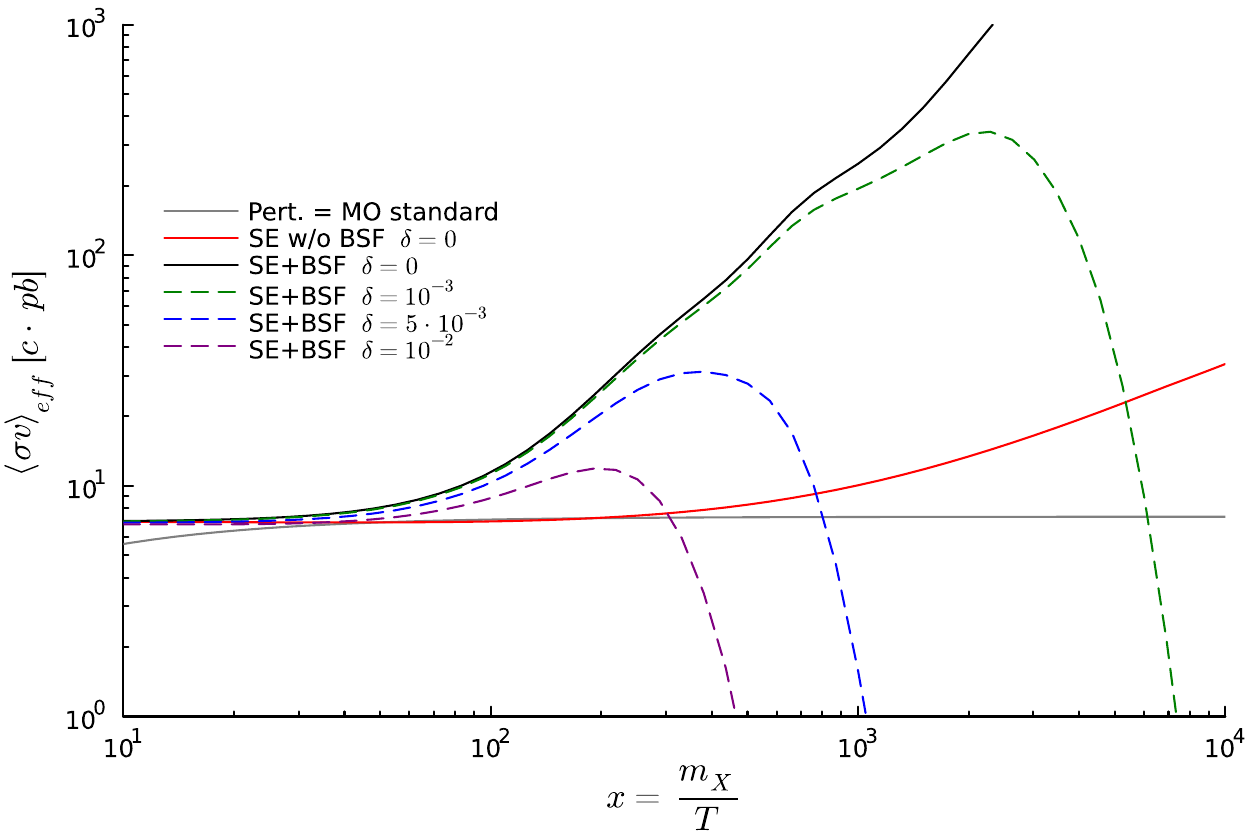}
    \caption{Dependence of the bound-state enhanced cross section on mass splitting $\delta$ in the \texttt{F3SuR} model for a DM mass $m_X = 1$\,TeV, a $t$-channel coupling $\lambda = 0.01$ and including SE and $n = 15$ bound states using the \textbf{Full Matrix} method. Smaller splittings allow longer-lived co-annihilation, enhancing late-time bound-state formation.}
    \label{fig:Cross_Sections_detail}
\end{figure}

Figure~\ref{fig:Cross_Sections} shows the different methods for obtaining the effective cross section for the \texttt{F3SuR} model with degenerate masses between DM and the Dirac mediator ($\delta = 0$, with $\delta \equiv \Delta m/m_X = (m_Y - m_X)/m_X$ the relative mass splitting in the dark sector).  
This degeneracy in mass is not a physically relevant benchmark point, but it is chosen because for $\delta=0$ the effective mediator cross section does not suffer from a suppression of the coannihilation weight in Eq.~\eqref{eq:effective_sigmav_coannih} (see also Figure~\ref{fig:Cross_Sections_detail}).
The gray baseline shows the purely perturbative cross section, while the orange curve includes SE.
The \textbf{Full matrix} solution, including $n=15$ bound state levels (solid red) and represents the most complete treatment.
For physical consistency, we restrict $n \leq 20$, ensuring the binding energy is always above the QCD scale for TeV-scale mediators\footnote{A related effect to confinement is the fact that for mass splittings as low as $\Delta m \lesssim \Lambda_\text{QCD}$, coannihilations of hadronic QCD bound states increase the effective annihilation cross section by a lot \cite{Gross:2018zha}. 
This is why the mass splitting should be chosen $\Delta m \gtrsim 1$\,GeV for consistency.}.
The dotted black \textbf{Ionization equilibrium} curve results in the largest effective annihilation cross section among the possible configurations/approximations. 
However, this limit is unphysical for large $x$ as the decay will dominate, which can be seen by comparing the black dotted line (ionization equilibrium) with the correct solution (red solid line) in Fig.\,\ref{fig:Cross_Sections}.
The \textbf{No transition} approximation (dashed black) provides excellent agreement with the full solution at freeze-out temperatures ($x \sim 20-30$) with significantly lower computational cost, while the \textbf{Efficient transition }method (dash-dotted black) overestimates the cross section.
It could be used to obtain conservative upper bounds on $\langle \sigma_\text{BSF} v \rangle_\text{eff}$.

The colored lines show the effective dark matter annihilation cross sections, taking into account the Sommerfeld effect and bound state formation with bound states with main quantum numbers $n=1-5$, which are calculated using the \textbf{Full matrix} solution for bound states.
At freeze-out, the ground state ($n=1$) dominates, while excited states become increasingly important at later times.
The dash-dot-dotted lines show spin-triplet contributions, which are suppressed at freeze-out by small decay widths (spin triplet states decay into three instead of two gluons) but dominate at low temperatures due to a statistical factor stemming from the spin sum in the cross section: The spin singlet cross section comes with a prefactor $\frac{1}{4}$, while the spin triplet cross section comes with a prefactor $\frac{3}{4}$. 

As shown in Figure~\ref{fig:Cross_Sections_detail}, the late-time enhancement from excited states ($n>1$) and spin-triplets is strongly suppressed by non-zero mass splitting $\delta > 0$ between dark matter and co-annihilating partners, as the effective cross section experiences an additional $e^{-2x\delta}$ suppression. Therefore bound-state effects are most pronounced for nearly degenerate spectra, making them particularly relevant for models with small mass splittings or models where the DM candidate itself is color-charged.

\section{Performance and Limitations}\label{sec:numerics}

\subsection{Performance}\label{subsec:Performance}
To quantify the computational cost of including non-perturbative effects, we conducted a series of performance benchmarks. 
All timing measurements were performed on a desktop workstation with an \texttt{Intel Core i7-12700K} processor and \texttt{16 GB} of RAM, running \texttt{Ubuntu 20.04 LTS}.
The code was compiled and executed using the standard \texttt{micrOMEGAs 6.2.4} build system.
Reported runtimes represent the average real time over 100 executions for each configuration. \\

The performance of \texttt{SE+BSF4DM} is highly model-dependent, influenced by factors such as mediator masses (relative to the DM mass), the number of mediator species, and their spins. For this study, we present benchmarks for the \texttt{F3SuR} model with a Dirac fermionic mediator and a mass splitting of $\delta = 10^{-2}$. Table~\ref{tab:performance_compact} summarizes the results. 
The columns correspond to the possible choices for \texttt{bsf\_scenario}, introduced in Sec.~\ref{subsec:code_description_BSF}.
The benchmark value is the runtime of a standard \texttt{micrOMEGAs} calculation with the \texttt{darkOmega} method, not making use of \texttt{SE+BSF4DM}. \\
Including the Sommerfeld effect alone (utilizing either the \texttt{darkOmega} or \texttt{darkOmegaExt} routines) introduces a negligible computational overhead compared to the perturbative calculation.
Adding the ground-state ($n=1$) bound state contribution also has a minimal impact.

However, the runtime increases significantly when including excited states ($n>1$), with the scaling behavior strongly dependent on the method used to solve the network of Boltzmann equations. The efficient transition limit (\textbf{Eff.~Tran}), where all $\ell$ bound state levels contribute, scales approximately as $\mathcal{O}(n^2)$.
The \textbf{Full Matrix} method, diagonalizing the matrix $M$ in Eq.~\eqref{eq:BSF_master_equation} by computing all QED mediated $\ell \rightarrow \ell'$ transitions, scales as $\mathcal{O}(n^4)$, becoming computationally intensive for large $n$. 
We restrict our analysis to $n \leq 20$, as for higher principal quantum numbers the binding energy $E_{\mathcal{B}_n}$ falls below $\Lambda\text{QCD}$ for dark matter masses below $\sim 10$ TeV, rendering the weakly coupled bound state picture invalid. The ionization equilibrium (\textbf{Ion.~Eq.}) approximation becomes unphysical at low temperatures, and the cross sections become excessively large (as theoretically expected).

We currently use \texttt{micrOMEGAs}' internal mass diagonalization routine \texttt{rDiagonalA} for computing $M^{-1}$ in Eq.~\eqref{eq:BSF_master_equation}, specifically by diagonalizing $M \rightarrow U^\dagger M_D U$ and computing its inverse via $M^{-1} = U M_D^{-1} U^\dagger$.
The calculation and subsequent inversion of the matrix $M$ represents the most computationally intensive part of this solution method at the moment. 

\begin{table}[h]
    \centering
    \caption{Performance benchmarks of the different methods used in \texttt{SE+BSF4DM} for the \texttt{F3SuR} model.}
    \label{tab:performance_compact}
    \small
    \begin{tabular}{@{}lccccc@{}}
        \toprule
        & \multicolumn{4}{c}{\textbf{Runtime (seconds)}} \\
        \cmidrule(l){2-6}
        \textbf{Process} & \textbf{No~BSF} \quad & \textbf{No~Tran.} \quad & \textbf{Eff.~Tran.} \quad & \textbf{Ion.~Eq.} \quad & \textbf{Full~Matrix} \\
        \midrule
        
        \textit{Baseline} \\
        Perturbative (\texttt{darkOmega}) & 0.04 & -- & -- & -- & -- \\
        SE only (\texttt{darkOmega}) & 0.05 & -- & -- & -- & -- \\
        Perturbative (\texttt{darkOmegaExt}) & 0.06 & -- & -- & -- & -- \\
        SE only (\texttt{darkOmegaExt}) & 0.07 & -- & -- & -- & -- \\
        SE + BSF $n=1$ (ground state) & -- & 0.07 & 0.07 & -- & 0.07 \\
        
        \midrule
        
        \textit{Excited states} \\
        SE + BSF $n=2$ & -- & 0.08 & 0.09 & 0.07 & 0.09 \\
         SE + BSF $n=3$ & -- & 0.10 & 0.11 & 0.07 & 0.11 \\
         SE + BSF $n=4$ & -- & 0.11 & 0.14 & 0.08 & 0.17 \\
         SE + BSF $n=5$ & -- & 0.12 & 0.20 & 0.08 & 0.26 \\
         SE + BSF $n=20$ & -- & 0.72 & 6.70 & -- & 98.00 \\
        \bottomrule
    \end{tabular}
\end{table}

In summary, while our implementation of the Sommerfeld effect introduces minimal overhead, the user's choice regarding the treatment of excited bound states is the primary determinant of computational cost.
As can be seen in Figure~\ref{fig:Cross_Sections}, the computationally fast no transition limit (\textbf{No~Tran.}) is a good approximation for the \textbf{Full Matrix} solution in the low $x$-region (which is the relevant one for coannihilations).

\subsection{Limitations and Future Developments}\label{subsec:extensions}
\texttt{SE+BSF4DM} is designed to be accessible and follows the logical structure of \texttt{micrOMEGAs}, ensuring ease of use. 

The most significant potential improvement concerns the treatment of conversion-driven freeze-out (coscattering) regimes of parameter space \cite{Garny:2017rxs, DAgnolo:2017dbv}, which exhibit a very small $t$-channel coupling $\lambda$ ($\mathcal{O}(10^{-9} - 10^{-7})$), spoiling chemical equilibrium in the dark sector. \\
The SE and BSF improved annihilation rates are implemented assuming chemical equilibrium and the standard coannihilation scenario to hold, where excited states provide only subleading corrections.
However, as demonstrated, excited states play a pivotal role in conversion-driven freeze-out \cite{Garny:2021qsr} and in the Super-WIMP scenario \cite{Binder:2023ckj}. \\ 
The implementation of this functionality is conceptually straightforward within our framework and would leverage \texttt{micrOMEGAs}' existing formalism for N-component dark matter \cite{Alguero:2023zol}, which can also be used to solve the dark matter time evolution in a scenario of conversion-driven freeze-out. 
This would merely require the future availability of a \texttt{darkOmegaNExt} function, which would extend the features of the currently available function \texttt{darkOmegaN}, which is an already available routine for multi-component dark matter models, by allowing to supplement it by an external thermally averaged cross section. 
We have been in contact with the \texttt{micrOMEGAs} developers and encourage its addition to \texttt{micrOMEGAs}. \\
Once available, combining \texttt{darkOmegaNExt} with \texttt{SE+BSF4DM} would enable automated relic density predictions in this regime, which is also experimentally interesting for long-lived particle searches at colliders \cite{Arina:2025zpi}.

Beyond extensions related to the relic density computation itself, the package is currently applicable if particles transform in the (anti)-fundamental representation of $SU(3)_C$.
Extending \texttt{SE+BSF4DM} to other representations, such as color-octet mediators, would allow to study a larger class of models, but also require more involved color decompositions for both the SE and BSF.



\section{Conclusions}\label{sec:conclusions}
This manual has described the installation, configuration, and usage of \texttt{SE+BSF4DM}, a package for \texttt{micrOMEGAs} that enables the calculation of the dark matter relic density including the Sommerfeld effect and bound-state formation for QCD-colored particles. 
The package requires minimal modification from the user and integrates seamlessly into the standard \texttt{micrOMEGAs} workflow without interfering with its other functionalities.
It is applicable to a class of well-motivated models, rendering it a versatile tool for dark matter phenomenology.

The main work \cite{Becker2026} explores the phenomenological impact of these effects, including a comprehensive analysis of the relic density and experimental constraints. 
Readers are referred to that work for physics results, detailed phenomenological analysis, and interpretation of the effects of SE and BSF on dark matter constraints.

A key assumption in the current implementation is the maintenance of chemical equilibrium within the dark sector.
As discussed, a particularly promising development is the treatment of conversion-driven freeze-out, which could be readily implemented upon the future availability of the proclaimed \texttt{darkOmegaNExt} function in a future release of \texttt{micrOMEGAs}. 

We hope that \texttt{SE+BSF4DM} will become a valuable tool for the community and that this manual facilitates its adoption by researchers studying dark sectors with QCD-charged particles.
The inclusion of Sommerfeld and bound-state effects provides more accurate predictions for relic density calculations, ultimately leading to more robust comparisons with experimental constraints.

\section*{Acknowledgements}
We thank Alexander Pukhov for his support during the implementation in \texttt{micrOMEGAs}, and Pablo Olgoso for feedback on the draft. 
M.~N thanks Merlin Reichard and Julian Mayer-Steudte for useful input on the code and Stefan Lederer for fruitful discussions.
M.~B., E.~C. and J.~H. knowledge support from the Emmy Noether grant “Baryogenesis, Dark Matter and Neutrinos: Comprehensive analyses and accurate methods in particle cosmology” (HA 8555/1-1, Project No. 400234416) funded by the Deutsche Forschungsgemeinschaft (DFG, German Research Foundation). M.~N. acknowledges support from the DFG Collaborative Research Centre “Neutrinos, Dark Matter and Messengers” (SFB 1258). M.~B., E.~C. and J.~H. acknowledge support by the Cluster of Excellence “Precision Physics, Fundamental Interactions, and Structure of Matter” ($\text{PRISMA}^+$ EXC 2118/1) funded by the DFG within the German Excellence Strategy (project No. 390831469).
This work was supported by Istituto Nazionale di Fisica Nucleare (INFN) through the Theoretical Astroparticle Physics (TAsP) project, and in part by the Italian MUR Departments of Excellence grant 2023-2027 “Quantum Frontiers”.
The work of M.B. was supported in part by the Italian Ministry of University and Research (MUR) through the PRIN 2022 project n. 20228WHTYC (CUP:I53C24002320006 and C53C24000760006). 

\bibliographystyle{apsrev4-2}
\bibliography{refs}

\end{document}